# 5-GHz Antisymmetric Mode Acoustic Delay Lines in Lithium Niobate Thin Film

Ruochen Lu, *Student Member*, Yansong Yang, *Student Member, IEEE*, Ming-Huang Li, *Member, IEEE*, Michael Breen, *Student Member, IEEE*, and Songbin Gong, *Senior Member, IEEE*

*Abstract*— We present the first group of acoustic delay lines (ADLs) at 5 GHz, using the first-order antisymmetric (A1) mode in Z-cut lithium niobate thin films. The demonstrated ADLs significantly surpass the operation frequencies of the prior art with similar feature sizes, because of their simultaneously fast phase velocity, large coupling coefficient, and low-loss. In this work, the propagation characteristics of the A1 mode in lithium niobate are analytically modeled and validated with finite element analysis. The design space of A1 ADLs is then investigated, including both the fundamental design parameters and those introduced from the practical implementation. The implemented ADLs at 5 GHz show a minimum insertion loss of 7.9 dB, an average IL of 9.1 dB, and a fractional bandwidth around 4%, with group delays ranging between 15 ns and 109 ns and the center frequencies between 4.5 GHz and 5.25 GHz. The propagation characteristics of A1 mode acoustic waves have also been extracted for the first time. The A1 ADL platform can potentially enable wide-band high-frequency passive signal processing functions for future 5G applications in the sub-6 GHz spectrum bands.

## I. INTRODUCTION

THE NEXT GENERATION radio access technology, namely the fifth-generation (5G) New Radio (NR), requires unprecedented signal processing capabilities [1], [2]. More specifically, the enhanced mobile broadband (eMBB), as one crucial 5G NR usage scenario targeting a thousand-fold increase in the mobile data volume per unit area [3], [4], is calling for novel wideband signal processing functions at radio frequency (RF). Acoustic signal processing, where the electromagnetic (EM) signals are converted and processed in the acoustic domain, is promising for providing chip-scale, low-loss, and wideband capabilities. First, acoustic devices feature miniature sizes because of the significantly shorter acoustic wavelengths ($\lambda$) compared to the EM counterparts, thus making them desirable for mobile applications [5], [6]. Second, designing and interconnecting acoustic devices can lead to passive implementation of signal processing functions [7], [8], which do not compete against the analog-to-digital converters (ADC) or digital signal processors (DSP) for the stringent power budget in RF frontends [9]. Last and most importantly, the recent demonstrations of low-loss and high electromechanical coupling ($k^2$) piezoelectric platforms [10]–[17] enable devices with lower insertion loss (IL) and wider fractional bandwidth (FBW), thus potentially overcoming the performance bottlenecks that currently hinder acoustic signal processing for eMBB applications.

Among various types of acoustic devices, acoustic delay lines (ADLs) have been demonstrated with diverse applications ranging from transversal filters [18], [19] and correlators [20]–[22] to oscillators [23], sensors [24], [25], and amplifiers [26]–[28], alongside the recent prototypes of time-domain equalizers [29] and time-varying non-reciprocal systems [30]. Conventionally, ADLs are built upon surface acoustic wave (SAW) platforms [31]–[33]. Despite their success in applications below 2 GHz, two main drawbacks hinder the broad adoption of SAW ADLs for eMBB applications. First, their moderate $k^2$ fundamentally limits the design trades in IL versus FBW [34]. In other words, it is challenging to achieve wide FBW without inducing substantial IL. Second, due to their slow phase velocity ($v_p$), it is challenging to scale the operation frequency above 3 GHz for the planned eMBB bands [1], [35], unless narrow electrodes (< 300 nm) [36], thin films on costly substrates [37]–[39], or intrinsically high damping modes [40] are adopted.

Recently, ADLs have been demonstrated with low loss and wide bandwidth using the fundamental shear horizontal (SH0) mode [41]–[43] and fundamental symmetrical (S0) mode [44]–[46] in suspended single-crystal lithium niobate (LiNbO₃) thin films enabled by the thin film integration techniques [47]. Compared with ADLs on other piezoelectric thin films [48]–[51], these demonstrations feature lower IL and larger FBW due to the simultaneously high $k^2$ and low damping of S0 [52], [53] and SH0 [54]–[56] modes in LiNbO₃. Nevertheless, it remains challenging to scale them above 3 GHz without resorting to narrow electrodes and ultra-thin films (<300 nm) [44], which are undesirable in terms of fabrication complexity and mostly lead to spurious modes that limit the achievable FBW [42]. Therefore, a new piezoelectric platform with simultaneously high $v_p$, large $k^2$, and low-loss is sought after for potential eMBB applications.

To this end, acoustic devices using the first-order antisymmetric (A1) mode in Z-cut LiNbO₃ have been reported with high $k^2$ and low loss above 4 GHz [57]–[59]. Different from SH0 and S0, A1 is higher-order in the thickness direction, thus significantly enhancing $v_p$ in in-plane dimensions [60] and improving frequency scalability. However, the highly dispersive

Manuscript submitted July 10th, 2019. This work was supported by the DARPA Microsystems Technology Office (MTO) Near Zero Power RF and Sensor Operations (N-ZERO) and Signal Processing at RF (SPAR) programs.

The authors are with the Department of Electrical and Computing Engineering, University of Illinois at Urbana-Champaign, Urbana, IL 61801 USA (email: rlu10@illinois.edu).

nature of A1 presents new challenges in designing ADLs. Design principles for S0, SH0, and SAW ADLs have to be revisited and substantially modified for A1 ADLs. Moreover, the notable cut-off in A1 confines acoustic waves within the input transducers and prevents their propagation toward the output port. Such effects are especially pronounced in the presence of metallic electrodes [60], and thus have to be analyzed and circumvented for successful implementation of A1 ADLs. Finally, A1 devices demonstrated so far are analyzed for mostly standing wave structures (e.g., resonators [60]) where the A1 propagation characteristics have not been systematically studied.

To overcome these outstanding hurdles, we aim to provide a comprehensive framework in this paper for analyzing the key parameters and propagation characteristics of A1 waves in LiNbO₃ thin films and subsequently implement wideband and high-frequency A1 ADLs. The fabricated ADLs show a minimum IL of 7.9 dB, an average IL of 9.1 dB, and a fractional bandwidth around 4%, delays ranging between 15 ns and 109 ns, and the center frequencies between 4.5 GHz and 5.25 GHz.

This paper is organized as follows. Section II provides a general discussion on the design of the A1 ADLs, focusing on A1 propagation characteristics and key parameters of A1 ADLs. Section III introduces the practical considerations essential for A1 ADLs, including electrode configurations and device orientations. Section IV presents the fabricated 5-GHz A1 ADLs. Section V presents the measured results of ADLs. A1 propagation characteristics, including the propagation loss (PL), $v_g$ and $v_p$, are also experimentally extracted. Finally, the conclusion is stated in Section VI.

## II. ASYMMETRIC MODE ACOUSTIC DELAY LINE

### A. Acoustic Delay Line Overview

The schematic of a typical A1 ADL is shown in Fig. 1 with the key parameters explained in Table I. The ADL consists of 30 nm thick aluminum interdigitated transducers (IDTs) on top of a suspended 490 nm Z-cut LiNbO₃ thin film. The thickness of LiNbO₃ is selected for enabling wideband operation at 5 GHz. More details will be provided in Section II-B. A pair of bi-directional transducers are placed on the opposite ends of the ADL. The transducers are composed of $N$ pairs of cascaded transducer unit cells. Each cell has a length of $\Lambda$, over which is situated a pair of transduction electrodes (each $\Lambda/4$ wide) with separations of $\Lambda/4$ in between. The electrodes are alternatively connected to signal (orange IDTs for Port 1, green IDTs for Port 2) and ground (blue IDTs). The in-plane orientation of the device is shown in Fig. 1, with the material's X-axis along the wave propagation direction (longitudinal direction). The orientation selection will be further discussed in Section III-B. Free boundaries, i.e., etch windows, are in the transverse direction for defining the acoustic waveguide. In operation, the RF signals are sent to Port 1 and converted into acoustic waves through the piezoelectric transducers. The launched acoustic waves propagate toward both ends, therefore sending half of the power toward Port 2. The other half is lost in the attenuation and scattering into the substrate. Similarly, after traversing through the waveguide with a gap length of $L_g$, only half of the power launched toward Port 2 is collected, causing a minimum IL of 6 dB. Various acoustic signal processing functions can be passively implemented through designing the transducers [7] and the waveguide [61]. The 6-dB IL from the bi-directional transducers can be effectively reduced using unidirectional transducers [18], [34], [46] with smaller feature sizes. In this work, we will focus on implementing the first group of A1 ADLs using bi-directional transducers.

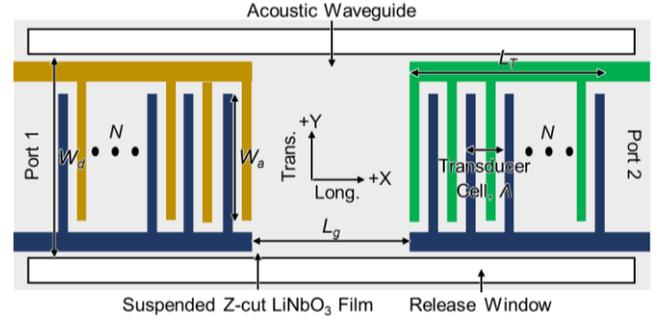

Fig. 1. Mockup of an A1 ADL on a suspended Z-cut LiNbO3 thin film.

TABLE I DESIGN PARAMETERS OF A1 ADLs

| Sym. | Parameter | Value | Sym. | Parameter | Value |
|------|-----------|-------|------|-----------|-------|
| $\Lambda$ | Cell length (μm) | 2.0-3.2 | $W_a$ | Aperture width (μm) | 50 |
| $N$ | Number of cells | 2-4 | $W_d$ | Device width (μm) | 74 |
| $L_g$ | Gap length (μm) | 20-320 | $L_T$ | Transducer length (μm) | 4.8-14.4 |
| $T_{LN}$ | LiNbO₃ thickness (nm) | 490 | $T_{Al}$ | Aluminum thickness (nm) | 30 |

### B. A1 Mode in Lithium Niobate Thin Film

Considering a piece of Z-cut LiNbO₃ waveguide (XZ plane) with infinite length in the Y direction, the wave propagation problem becomes two-dimensional (2D). Because of the planar geometry, the transverse resonance method [62] is used to solve the 2D vibration. In such a method, the modal solutions are decomposed into the traveling waves along the waveguide direction and the resonant standing waves in the transverse direction. The approach has been proven for both the acoustic and the EM cases [62], [63]. For a lossless and isotropic plate with mechanically free boundary conditions on the top and bottom surfaces, the symmetric and antisymmetric solutions can be analytically expressed using the Rayleigh-Lamb frequency equations [62]:

$$\frac{\tan(k_{ts} \cdot b/2)}{\tan(k_{tl} \cdot b/2)} = -\left(\frac{4\beta^2 k_{ts} k_{tl}}{(k_{ts}^2 - \beta^2)^2}\right)^{\pm 1} \quad (1)$$

$$k_{tl}^2 = (\omega/v_l)^2 - \beta^2 \quad (2)$$

$$k_{ts}^2 = (\omega/v_s)^2 - \beta^2 \quad (3)$$

where $k_{tl}$ and $k_{ts}$ are the transverse wavenumbers for the longitudinal and shear modes. $b$ is the film thickness, $\beta$ is the lateral wavenumber, and $\omega$ is the angular frequency. $v_l$ and $v_s$ are the velocities of the longitudinal and shear modes. In equation (1), the "+" and "−" are used to denote Lamb wave solutions of symmetrical and antisymmetric modes, respectively. Note that equations (1)-(3) are more complex than those for a rectangular EM waveguide because the longitudinal and shear acoustic

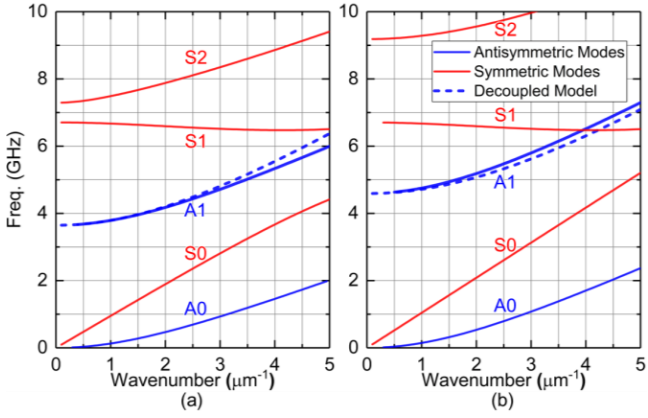

Fig. 2. Solutions of Lamb waves in a 490 nm Z-cut LiNbO₃ thin film under the isotropic and quasi-static approximations. (a) Electrically short and (b) electrically open boundary conditions. Different modes are marked on their solutions along with the ones from the decoupled model for A1 (dashed blue lines).

waves co-exist in the waveguide and mode conversion happens at the top and bottom surfaces [62].

Although solutions for Lamb waves in isotropic media can be solved using equations (1)-(3), the solutions in anisotropic piezoelectric thin films (e.g., LiNbO₃) are difficult to attain analytically unless certain acoustic modes along particular crystal orientations are studied [64], [65]. Finite element analysis (FEA) is one alternative for solutions. However, it does not provide straightforward insights into the principles of A1 propagation. To this end, we first introduce two approximations for a simplified model. The first one is the isotropic assumption in which the in-plane and out-of-plane stiffness constants are deemed the same for LiNbO₃. The second assumption is the quasi-static approximation [62], in which the electric field is assumed to have zero curl. Therefore, $v_l$ and $v_s$ in a plate with electrically short boundary conditions on both top and bottom surfaces can be approximated by:

$$v_l \approx \sqrt{c_{11}^E / \rho} \qquad (4)$$

$$v_s \approx \sqrt{c_{44}^E / \rho} \qquad (5)$$

where $c_{11}^E$ and $c_{44}^E$ are stiffness constants related to the longitudinal and shear waves respectively, following the Voigt notation [62], and $\rho$ is the material density. For single-crystal LiNbO₃, $c_{11}^E$ is $2.03 \times 10^{11}$ N/m², $c_{44}^E$ is $0.60 \times 10^{11}$ N/m², and $\rho$ is 4700 kg/m³ [66]. By solving equations (1)-(5) for $t$=490 nm, the estimated Lamb wave dispersion curves are attained and plotted in Fig. 2 (a) for the electrically short case. The A1 mode of interest is the second group of antisymmetric solutions, which are at higher frequencies than the fundamental antisymmetric mode (A0) mode with the same $\beta$. A1 exhibits a cut-off frequency, below which A1 waves do not have purely real $\beta$. In other words, only evanescent A1 waves, which attenuate exponentially with distance, exist below $f_{c\_short}$ in LiNbO₃ with the electrically short surface.

Similarly, the dispersion curves in a piezoelectric slab with electrically open boundary conditions can be calculated using the piezoelectrically stiffened elastic constants $c_{ij}'$, as [66]:

$$c_{ij}' = \left[c_{ijkl}^E + (e_{pij}e_{qkl}n_p n_q)/(\varepsilon_{rs}^S n_r n_s)\right] n_i n_l \qquad (6)$$

where $i$, $j$, $k$, $l$, $p$, $q$ are the indices of the Cartesian coordinate system, $\boldsymbol{n}$ is the unit vector, and $e$, $\varepsilon^S$ are the piezoelectric and dielectric constants, respectively. Equation (6) describes that the material stiffening due to the piezoelectric effect depends on the piezoelectric constants. For LiNbO₃, $c_{11}'$ is $2.19 \times 10^{11}$ N/m², and $c_{44}'$ is $0.95 \times 10^{11}$ N/m² [66]. By replacing the corresponding $c^E$ with $c'$ in equations (4)-(5), Lamb wave dispersion curves are attained and plotted in Fig. 2 (b) for the electrically open case. Likewise, a cut-off frequency $f_{c\_open}$ can be observed. For a given $\beta$, A1 is at higher frequencies compared to the previous case [Fig. 2 (a)] due to stiffening.

Equations (1)-(6) are still cumbersome for follow-on analysis of A1 ADLs. Therefore, we introduce the last assumption to decouple longitudinal and shear waves in A1 [60]. The dispersion of A1 can then be approximated by:

$$\omega^2 = (2\pi f_c)^2 + \beta^2 \cdot v_l^2 \text{ or } f^2 = f_c^2 + v_l^2/\lambda^2 \qquad (7)$$

$$f_c = v_s/(2b) \qquad (8)$$

where $f$ is the frequency, $\lambda$ is the wavelength, and $v_{l\_short}$ and $v_{l\_open}$ are the longitudinal wave velocities of respective cases. For a 490 nm LiNbO₃ thin film, $f_{c\_short}$ is 3.64 GHz, $v_{l\_short}$ is 6572 m/s, $f_{c\_open}$ is 4.59 GHz, and $v_{l\_open}$ is 6795 m/s [66]. The dispersion curves are plotted in Fig. 2 and compared with the results attained without the last assumption. The good agreement indicates that the model is adequate for A1 at small thickness-wavelength ratios ($h/\lambda$) [62].

From equations (7)-(8), it is clear that the film thickness $b$ determines the dispersion of A1. For a 5 GHz center frequency, $b$ has to be neither too small (450 nm for $f_{c\_open}$ at 5 GHz) to avoid the cut-off, nor too large (670 nm for a $\lambda$ of 1.6 μm at 5 GHz for electrically short) to avert small feature sizes. Thus, 490 nm is chosen as a trade-off. More discussion in the context of ADL designs will be presented in Section II-C.

To validate the simplified model and obtain more accurate properties of A1, eigenmode FEA is set up in COMSOL for a 490 nm Z-cut LiNbO₃ thin film section with a width (the +X direction) of $\lambda$. Periodic boundary conditions are applied to the XZ and YZ planes in both the electrical and mechanical domains. The top and bottom surfaces (XY planes) are set to be mechanically free. The electrical boundary conditions are set to be electrically open and short, respectively [44]. The simulated A1 dispersion curves (with different $\beta$) are presented in Fig. 3 (a). Similar to the analytical model, the cut-off phenomenon is also seen, showing an $f_{c\_short}$ of 3.66 GHz and an $f_{c\_open}$ of 4.37 GHz. The eigenfrequency increases for a larger $\beta$, suggesting that the center frequency ($f_{center}$) of A1 devices can be tuned by changing $\lambda$. More specifically, one can tune $f_{center}$ from 4.5 to 6.0 GHz by changing $\lambda$ from 6 μm to 1.5 μm for the electrically open case. In comparison, the simplified model provides a good estimation of A1 properties without resorting to the time-consuming calculation. Therefore, the model will be used in the later subsections for analyzing the A1 ADL design

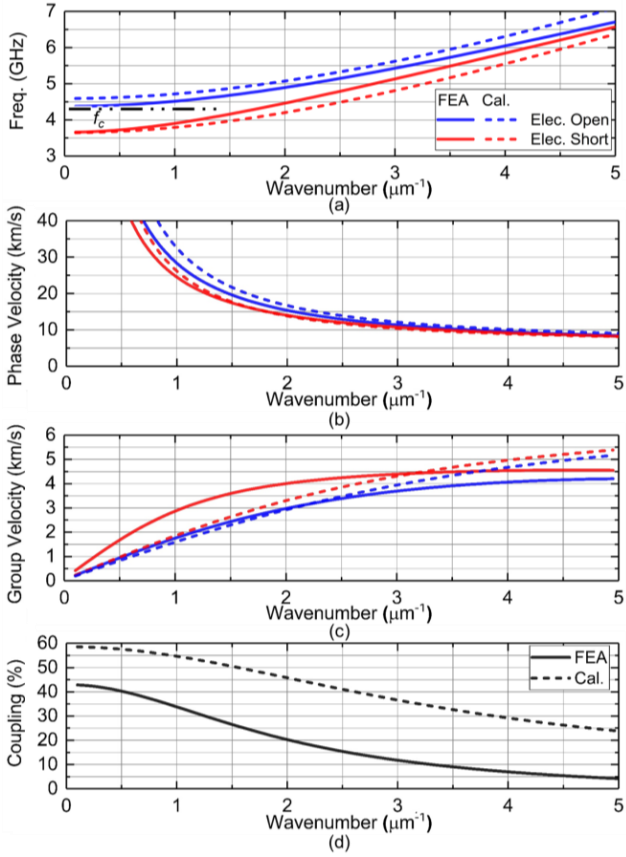

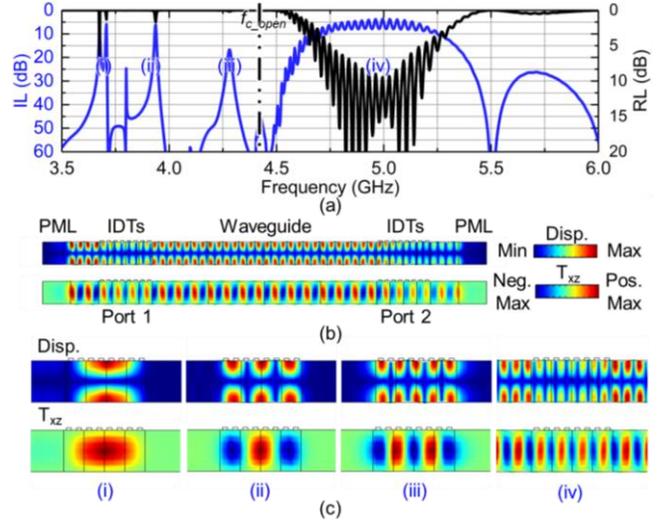

Fig. 4. (a) Simulated IL and RL with both ports conjugately matched. The evanescent modes in the input transducers are labeled. (b) Displacement and $T_{xz}$ stress distribution at the center frequency. (c) Displacement mode shapes and stress distributions in the input transducers at the marked frequencies.

Fig. 3. Characteristics of A1 with different $\beta$ in a 0.49 μm thick Z-cut LiNbO3 thin film, obtained from both FEA and the simplified model. (a) Eigen frequency, (b) $v_p$, and (c) $v_g$ with electrically open and short boundary conditions. (d) $k^2$ at different wavenumber.

Moreover, based on the eigenmode analysis, the phase velocity $v_p$ and the group velocity $v_g$ are [62]:

$$v_p = \omega/\beta \approx \sqrt{(2\pi f_c/\beta)^2 + v_l^2} \qquad (9)$$

$$v_g = \partial\omega/\partial\beta \approx v_l^2/\sqrt{v_l^2 + (2\pi f_c/\beta)^2} \qquad (10)$$

The obtained values are plotted in Fig. 3 (b) and (c), respectively. A remarkably high $v_p$ over 8000 m/s is obtained for A1 below 6.5 GHz. A low $v_g$ below 4500 m/s is also observed. Moreover, the mode is highly dispersive, and thus requires careful design for the targeted operation frequency. $v_p$ and $v_g$ calculated from the simplified model using equations (9)-(10) are also plotted, matching the trend of the simulated values.

$k^2$ is then calculated from $v_p$ by [67], [68]:

$$k^2 = (v_f^2 - v_m^2)/v_m^2 \qquad (11)$$

where $v_f$ and $v_m$ are the phase velocities of the electrically open and short cases. The dispersion curve of $k^2$ is plotted in Fig. 3 (d). High $k^2$ above 40% can be observed for A1 waves with a long $\lambda$ (or with operation frequencies close to $f_c$). $k^2$ declines for A1 waves at a higher frequency (or with a larger $h/\lambda$). Nevertheless, $k^2$ larger than 10% is obtained for 5.5 GHz devices ($\lambda$ of 2 μm). The discrepancies between the simulated and the calculated $k^2$ are due to the assumption in the simplified model, which induces overestimation in the frequency difference between $v_f$ and $v_m$, and thus causes a larger calculated $k^2$.

With the critical characteristics of A1 studied, it is apparent that A1 ADLs are promising for 5G applications for several reasons. First, a high $v_p$ enables high-frequency devices without resorting to narrow electrodes or thin films [44]. Based on Fig. 3, it is feasible to achieve 5 GHz with a 600 nm feature size on 490 nm thick Z-cut LiNbO3 [58]. Second, the slow $v_g$ of A1 (e.g., 3000 m/s at 5 GHz) enables longer delays over the same length in comparison to alternatives with faster $v_g$ (e.g., S0, or SH0) [62], thus permitting a smaller device footprint. Third, large $k^2$ above 5 GHz can overcome conventionally unforgiving trades between IL and FBW [34], consequently allowing low-loss and wide-band signal processing functions. For example, up to 30% FBW is accessible without significantly increasing IL at 5.5 GHz ($k^2$ of 15%) [42].

### C. Simulation of A1 Acoustic Delay Line

The typical response of an A1 ADL will be first studied using 2D FEA. The 2D FEA assumes that the acoustic waves are plane waves propagating along the X-axis (the longitudinal direction in Fig. 1), neglecting the fridge effects near the release windows. The three-dimensional (3D) case will be presented in Section III-B, emphasizing the in-plane propagation characteristics. As presented in [44], perfectly matched layers (PML) are applied to the longitudinal ends of the ADL. The simulation assumes lossless conditions in both the electrical and mechanical domains because the loss factors in LiNbO3 thin films at RF are currently not well understood and remain an active area of experimental research [44], [69]. Note that massless electrodes are used in this section for simplicity. The effects of mechanical loading from practical electrodes will be presented in Section III-A.

An A1 ADL prototype (cell length $\Lambda = 2.4$ μm, gap length $L_g = 40$ μm, and cell number $N = 4$) is simulated to showcase its typical frequency domain response (Fig. 4). The aperture width of the device (transverse direction, along the Y-axis in Fig. 1) is set as 50 μm. The S-parameters are obtained from the frequency domain FEA and then conjugately matched with $360 + j30 \ \Omega$ for both the input and output ports [Fig. 4 (a)], showing a well-defined passband centered around 5 GHz. Note that, although a complex port impedance is used in the example, it is possible to achieve acceptable matching over a large FBW using a real port impedance without significantly increasing IL, thanks to the large $k^2$ of A1 [34]. Such a high operation frequency is as predicted in the eigenmode analysis, validating the choice of 490 nm-thick LiNbO₃. The displacement mode shape and the stress distribution ($T_{xz}$) at the center frequency are plotted in Fig. 4 (b). The minimum in-band IL is 3.7 dB, the average IL is 6.0 dB, and the 3-dB FBW is 10%. The average IL and 3-dB FBW reported in Sections II and III are extracted from the transmission after smoothing with a 50-point-window in the frequency domain using the Savitzky-Golay approach [70]. The 6-dB IL is caused by the bi-directional loss. The slight ripples in RL and IL are caused by triple transit signals (TTS) between the input and output transducers, which are intrinsic to ADLs employing bi-directional transducers [34].

Different from S0 and SH0 ADLs, the A1 ADL features a non-symmetric passband, which is apparent from the sidelobes. The non-symmetry is caused by the cut-off of A1 (cut-off frequency of the LiNbO₃ thin film with electrically open surfaces, $f_{c\_open}$ labeled in Fig. 4). As explained in Section III-A, A1 waves at frequencies below $f_{c\_open}$ are evanescent. Thus, the amplitude decays during the propagation toward the output transducers. Below $f_{c\_open}$, the section with the input transducer is equivalent to an A1 mode resonator [57]–[60]. The acoustic impedance difference caused by different electrical boundary conditions acts as reflective boundaries [7], [41]. The resonant modes below $f_{c\_open}$ are marked with (i)-(iii) in the frequency response [Fig. 4 (a)]. Their displacement and stress mode shapes are shown in Fig. 4 (c). Only odd lateral order A1 resonances are built up in the input transducer because the charge generated from even-order lateral overtones is canceled in a 4-cell transducer. At odd mode resonances, a small portion of the energy build-up in the input transducers leaks to the output through evanescent coupling. Therefore, resonances in IL and RL are also seen at these frequencies. These modes are only prominent in the simulation because the structure is set as lossless. It can be seen in   Section V that they are significantly attenuated in measurements. Naturally, we will focus on the frequency range above $f_{c\_open}$ to demonstrate A1 ADLs.

### D. Key Design Parameters of A1 Acoustic Delay Lines

In this subsection, the dependence of the three main ADL specifications, namely the group delay ($\delta$), center frequency ($f_{center}$), and FBW, on the device dimensions will be investigated.

The impact of $L_g$ on the obtained $\delta$ is first studied. FEA simulated IL, RL, and $\delta$ of ADLs with $L_g$ of 20, 40, and 80 μm are shown in Fig. 5, with ports matched to $360 + j30 \ \Omega$. For this group of devices, the average IL is 6.0 dB, and the 3-dB FBW is 10%. The results underline three key insights. First, $\delta$ increases in a highly dispersive fashion for devices with longer gaps. $v_g$ at each frequency is extracted using least square fitting [71]. The extracted $v_g$ is compared with that obtained from eigenmode simulations, showing good agreement [Fig. 5 (d)]. Such a dispersive delay can be advantageous for chirp compressors [72]. If a constant delay is required, one might inversely chirp $\Lambda$ of different cells in both transducers to compensate for the dispersion in $v_g$. Second, the periodicity of the ripples in the S-parameters is inversely proportional to the gap length, similar to that in S0 [44]. It shows that the ripples are caused by the reflections between transducers, which form a weak resonant structure. Last, the transmission of the modes below the cut-off frequency $f_{c\_open}$ decreases for longer devices. This verifies the evanescent nature of these modes, as suggested by our simplified model.

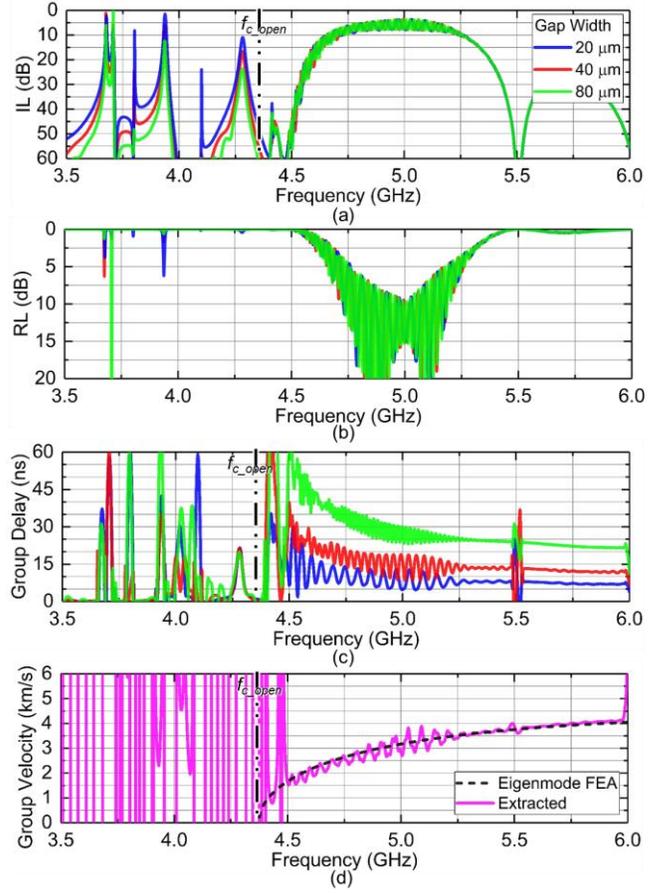

Fig. 5. Simulated (a) IL, (b) RL, and (c) group delay of A1 ADLs with different gap lengths. Different devices have the same cell number $N$ of 4, the same cell length $\Lambda$ of 2.4 μm, but different gap lengths $L_g$ of 20, 40, and 80 μm. (d) Extracted group velocity in comparison with that directly obtained from the eigenmode simulation.

The effects of $\Lambda$ on the center frequency $f_{center}$ are then investigated. $f_{center}$ is the frequency at which most RF energy is converted into the EM domain. FEA simulated IL, RL, and $\delta$ of ADLs with different $\Lambda$ of 2.4, 3.2, and 4.0 μm are shown in Fig. 6. The ports are matched to $360 + j30 \ \Omega$, $300 - j60 \ \Omega$, and $400 - j80 \ \Omega$, respectively. The average IL is 6 dB, 6 dB, and 11.7 dB, while the FBW is 10%, 6.8%, and 1.0%, respectively. The most apparent difference lies in $f_{center}$ and the passband shape. The effects can be explained using the Berlincourt equation. At $f_{center}$, the acoustic wavelength matches the transducer cell

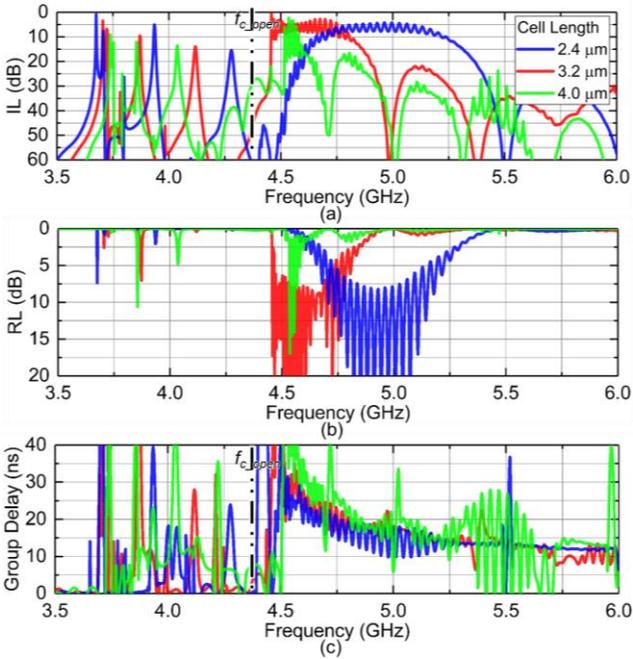

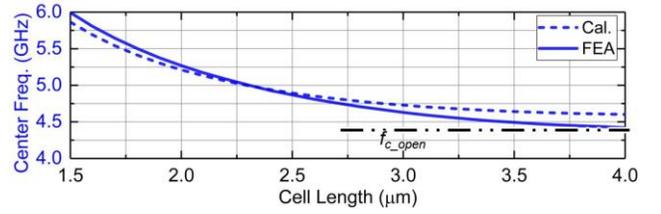

Fig. 7. Dependency of $f_{center}$ on the $\Lambda$.

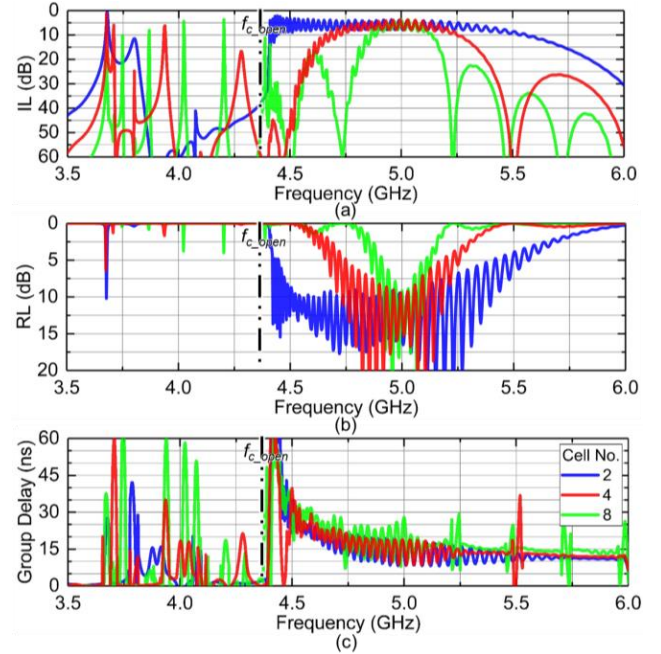

Fig. 6. Simulated (a) IL, (b) RL, and (c) group delays of the A1 ADLs with different center frequencies. Different devices have the same cell number $N$ of 4, and the same gap length $L_g$ of 40 μm, but different cell lengths $\Lambda$ of 2.4, 3.2, and 4.0 μm.

Fig. 8. Simulated (a) IL, (b) RL, and (c) group delays of the A1 ADLs with different bandwidths. Different devices have the same cell length $\Lambda$ of 2.4 μm, and the same gap length $L_g$ of 40 μm, but different cell number $N$ of 2, 4, and 8.

length [56]. Therefore, the equation for solving $f_{center}$ is:

$$f_{center} \cdot L_{open}/v_{p\_open} + f_{center} \cdot L_{short}/v_{p\_short} = 1 \qquad (12)$$

where $L_{open}$ and $L_{short}$ are the lengths of the parts without and with electrodes in a cell. $v_f$ and $v_m$ are the phase velocities in that area with electrodes (electrically short) and without electrodes (electrically open) respectively, which can be related to $f_{center}$ by a variation of equation (9) [62]:

$$v_f = v_{l\_open}/\sqrt{1-\left(f_{c\_open}/f\right)^2} \qquad (13)$$

$$v_m = v_{l\_short}/\sqrt{1-\left(f_{c\_short}/f\right)^2} \qquad (14)$$

Based on $f_c$ and $v_l$ calculated in Section II-B, $f_{center}$ for a 50% duty-cycled transducer with $\Lambda$ between 1.5 and 4.0 μm is shown in Fig. 7 as the blue dash line. $f_{center}$ keeps decreasing for an ADL with a larger $\Lambda$. However, as $f_{center}$ gets closer to $f_{c\_open}$, the passband is truncated and distorted, leading to a reduction of FBW. To validate the simplified model, FEA is used to validate the case. 4 pairs of transducers are simulated in the frequency domain. $f_{center}$ and the wavelength are plotted in Fig. 7. The simplified model agrees well with the simulation. In addition to the change in the passband, longer $\Lambda$ also lowers the frequencies of the non-propagating modes within the input transducers due to the longer resonant cavity.

Finally, the effects of $N$ on FBW are studied. FEA results of ADLs with different $N$ values of 2, 4, and 8 are shown in Fig. 8, with ports conjugately matched to $800 + j910$ Ω, $360 + j30$ Ω, and $112 - j80$ Ω. The average IL is around 6 dB, while the FBW is 21%, 10%, and 4.8%, respectively. The FBW of ADLs is roughly inversely proportional to the number of cells, as ex-plained by the transfer function of the transducer pair [7]. However, because of the cut-off phenomenon, the passband gradually distorts near $f_{c\_open}$. Therefore, one needs to consider thoroughly $f_{center}$ and the FBW requirements before designing A1 ADLs.

To sum up, we have discussed the principles and critical parameters ($\Lambda$, $L_g$, and $N$) of A1 ADLs. The discussions focus on ideal A1 ADLs without considering the mass loading of the electrodes. Furthermore, the actual aperture width and possible skewed propagation of A1 in a 3D structure are not captured by the adopted 2D simulations. The electrical loading in transducers is also ignored. All these practical considerations will be covered in Section III.

## III. IMPLEMENTATION CONSIDERATIONS

### A. Electrode Mass Loading

In this subsection, we will show the simulated results of ADLs using electrodes of different thicknesses and different metals. Different devices studied herein have the same cell length $\Lambda$ of 2.4 μm, gap length $L_g$ of 40 μm, and cell number $N$ of 4.

As seen in Fig. 9, the thickness of the electrode layer affects the performance. The S parameters for devices with electrodes of 0, 30, and 60 nm Al are conjugately matched with $360 + j30$ $\Omega$, $225 + j130\ \Omega$, and $145 + j220\ \Omega$ respectively. The drifting of $f_{center}$ to lower frequencies and slightly larger ripples are observed for ADLs with thicker metal. The influence of different types of metal (Al, Mo, and Au) with the same electrode thickness (30 nm) on the ADL performance is shown in Fig. 10. The results are matched with $225 + j130\ \Omega$, $105 + j215\ \Omega$, and $87 + j45\ \Omega$, respectively. The same trend in thicker electrodes can be observed for heavier films.

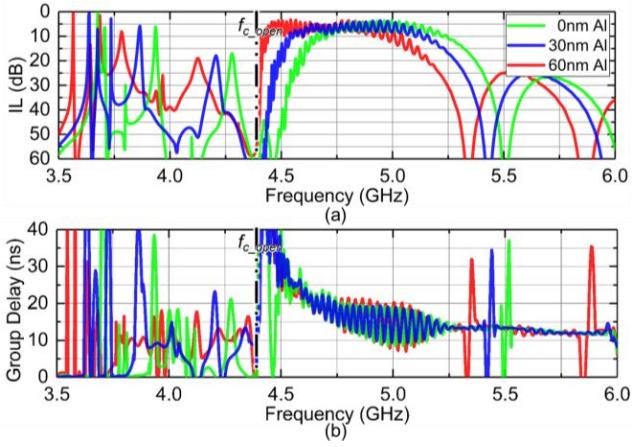

Fig. 9. Simulated (a) IL and (b) group delay of the A1 ADLs with aluminum electrodes of 0, 30, and 60 nm in thickness.

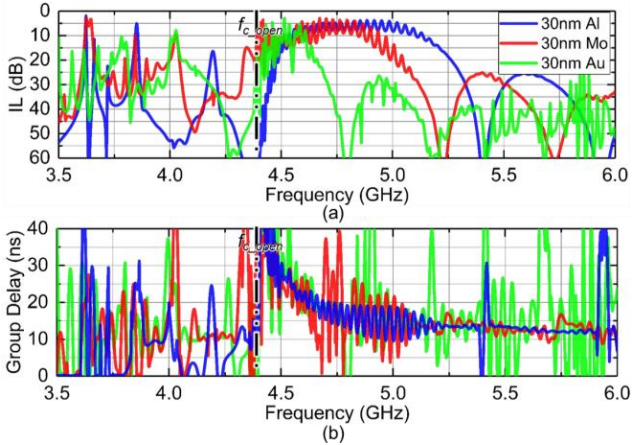

Fig. 10. Simulated (a) IL and (b) group delay of the A1 ADLs with 30 nm electrodes of different types of metal.

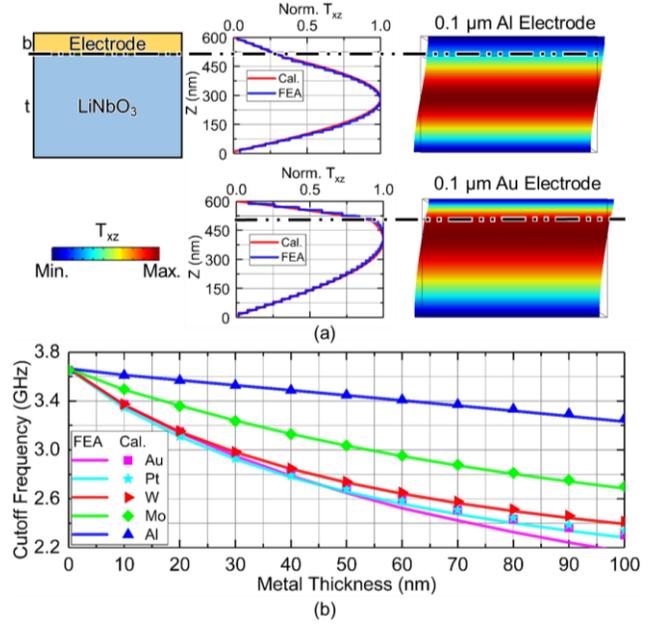

Fig. 11. (a) Stress ($T_{xz}$) distribution of A1 at $f_{c\_short}$ in film stacks with 490 nm LiNbO₃ and 100 nm metal on the top. The calculated stress distribution and that obtained from FEA are presented. (b) Calculated and simulated $f_{c\_short}$ for 490 nm LiNbO₃ and metal on the top.

To better design electrodes for A1 ADLs, the lower $f_{center}$ caused by the mass loading is first discussed. As presented in equations (12)-(14), $f_{center}$ is determined by the $v_l$ and $f_c$ in the parts with and without electrodes. For devices with different electrodes, both $f_{c\_short}$ and $v_{l\_short}$ vary.

First, $f_{c\_short}$ of different film stacks can be obtained analytically by solving the stress distribution ($T_{xz}$) in the film stack [Fig. 11 (a)]. At $f_{c\_short}$, $T_{xz}$ is uniform in the lateral direction. Given that the stress vanishes on the top and bottom surfaces with the mechanically free boundary conditions, the stress distribution can be described in the thickness direction ($z$) as:

$$T_{xz}(z) = T_{LN} \cdot \sin(2\pi f_{c\_short}/v_{s\_LN} \cdot z), \text{when } 0 \leq z < t$$
$$T_{xz}(z) = T_{met} \cdot \sin[2\pi f_{c\_short}/v_{s\_met} \cdot (t + b - z)], \quad (15)$$
$$\text{when } t \leq z \leq t + b$$

where $T_{LN}$ and $T_{met}$ are the stress amplitude, while $v_{s\_LN}$ and $v_{s\_met}$ are the shear wave velocities in LiNbO₃ and the electrode respectively. $t$ and $b$ are the thicknesses of LiNbO₃ and the electrode. Using the boundary conditions at the interface, namely the stress continuity and velocity continuity [8], [56], we have:

$$\frac{\tan(2\pi f_{c\_short}/v_{s\_met} \cdot b)}{\tan(2\pi f_{c\_short}/v_{s\_LN} \cdot t)} = -\frac{\rho_{LN} \cdot v_{s\_LN}}{\rho_{met} \cdot v_{s\_met}} \quad (16)$$

where $\rho_{LN}$ and $\rho_{met}$ are the densities of respective materials. $f_{c\_short}$ and the normalized stress distribution in the film can be obtained from equations (15)-(16). Two examples, 100 nm Al and 100 nm Au on the top of 490 nm LiNbO₃, are presented in Fig. 11 (a) for displaying the effect of mass loading on $f_{c\_short}$. The solutions for both cases are plotted, showing that the metal layer changes the stress distribution and consequently lowers $f_{c\_short}$. In the Au case, nearly half of the stress variance is in Au due to the significantly slower shear wave velocity in the metal layer. In contrast, the impact is much smaller in the Al case because of a faster shear wave velocity in Al. The mass loading effects caused by different metals are then calculated [Fig. 11 (b)]. Thicker electrodes and metals with slower $v_{s\_met}$ lead to a larger difference. Eigenmode FEA (Fig. 11) also shows great agreement with our analytical model.

Second, the $v_{l\_short}$ of different film stacks are solved through FEA. Although analytically solving a composite structure is possible through simplifications [73], FEA is used here for higher accuracy [Fig. 12 (a)]. 100 nm of Al leads to 1.1% velocity change, while 100 nm Au leads to 22.7% velocity change. Thicker or heavier electrodes lead to a more significant phase velocity decrease.

With the dependence of $f_{c\_short}$ and $v_{l\_short}$ on the electrode thickness studied, the impact of the mass loading on $f_{center}$ is calculated using the model in equations (12)-(14) and plotted in Fig. 12 (b) for a 50% duty cycle transducer with $\Lambda$ of 2.4 µm. Because both parameters decrease with thicker or heavier metal (e.g., Au), $f_{center}$ decreases, compared to the massless case. As a result, the passband distorts as it shifts to lower frequencies and gets truncated by the cut-off (Figs. 9-10). To build A1 ADLs at similar frequencies using thicker or heavy electrodes, one has to implement transducers with a smaller $\Lambda$, which requires a smaller feature size. Therefore, thin electrodes with fast wave velocities are preferred for achieving well-defined passbands without decreasing the feature size of IDTs.

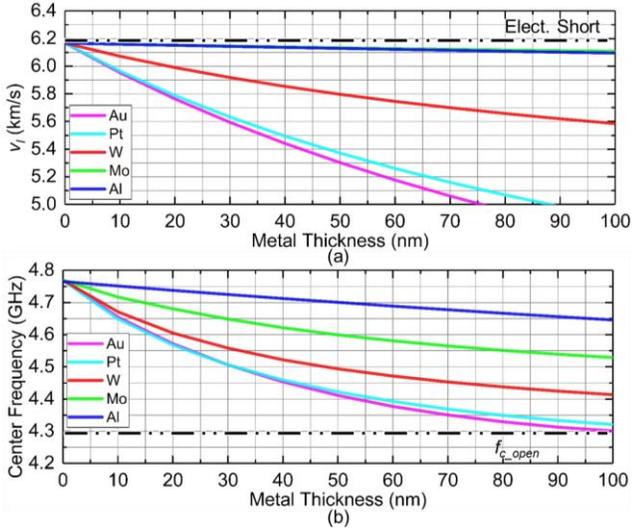

Fig. 12. (a) Simulated $v_{l\_short}$ in film stacks with 490 nm LiNbO₃ and different types of electrode on the top. (b) Calculated $f_{center}$ for transducers with different types of electrodes.

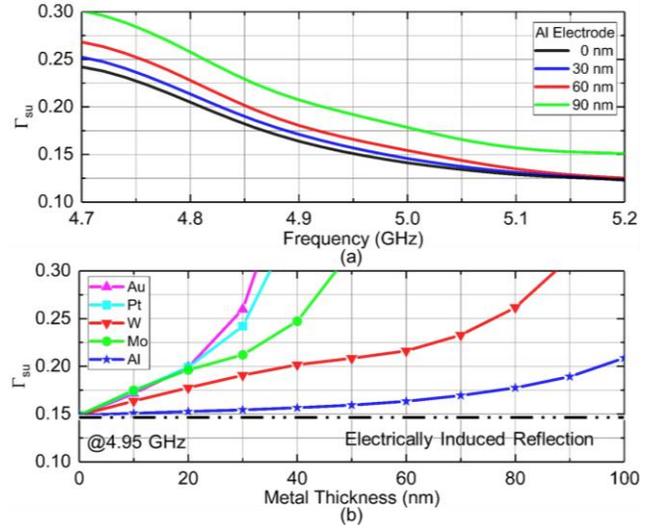

Fig. 13. (a) Simulated $\Gamma_{su}$ at the interface between the parts with and without electrodes. (b) Comparison of the $\Gamma_{su}$ caused by different metal electrodes for A1 waves in the acoustic waveguide at 4.95 GHz.

Another effect from more severe mass loading is the larger ripples in IL and group delay. These are caused by more significant internal reflections at the edge of the electrodes [34], [41]. The edge reflections are of two origins, namely the electrically induced $\Gamma_e$ and the mechanically induced $\Gamma_m$ [41]. While $\Gamma_e$ does not change with electrode thickness, $\Gamma_m$ is larger for thicker metals. To study the influence quantitatively, the reflection generated at the interface between the parts with and without electrodes, namely the step-up reflection coefficient, $\Gamma_{su}$, is studied using the simulation procedure in [41], [61]. A slab of LiNbO₃ partially covered with metal is modeled in 2D with PMLs on the lateral ends for absorbing the reflected waves. A1 waves are excited mechanically in the area without electrodes and propagate toward the interface. The ratio between the reflected stress and the incident stress ($T_{xz}$) is used to calculate $\Gamma_{su}$. $\Gamma_{su}$ for Al electrodes of different thicknesses is plotted in Fig. 13 (a). First, $\Gamma_e$ shows lower values at higher frequencies, which is consistent with the lower $k^2$ at these frequencies (Fig. 3). Second, larger $\Gamma_{su}$ is observed for thicker electrodes due to the larger mechanically induced reflections. The larger reflections subsequently induce larger in-band ripples, which are more severe near $f_{c\_short}$. Similarly, heavier material leads to larger reflections [Fig. 13 (b)]. Thus, a lighter electrode material such as Al is preferred to reduce $\Gamma_{su}$ for less pronounced in-band ripples [7], [18]. Note that the internal reflections can be further suppressed by split electrode designs at the cost of needing a smaller feature size [7].

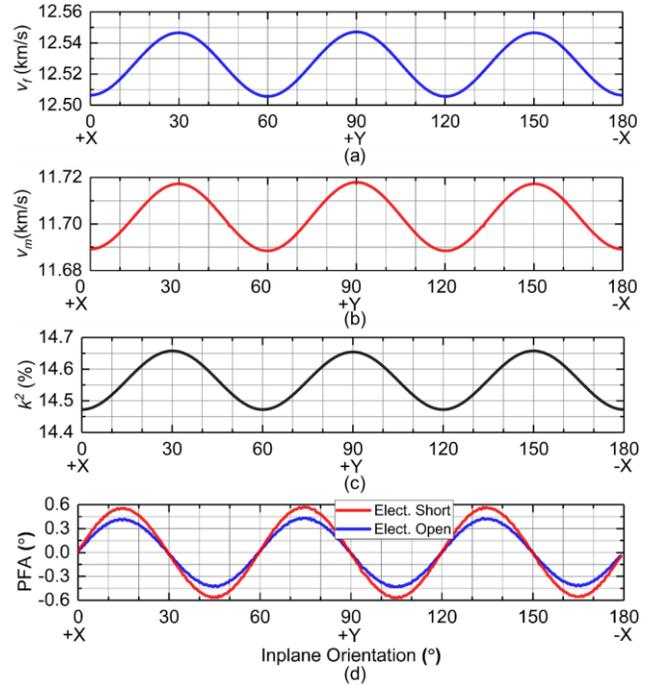

Fig. 14. Simulated A1 characteristics at different in-plane orientations in a 0.49 µm thick Z-cut LiNbO₃ thin film. Simulated $v_p$ under (a) electrically open and (b) short boundary conditions, (c) $k^2$, and (d) power flow angle.

To sum up, thinner electrodes with faster phase velocities are preferred in maintaining high-frequency and wide-band performance. However, if the electrodes are too thin, the series resistance would load the performance electrically (Section III-C). In our work, 30 nm Al is chosen as a calculated trade-off.

### B. In-plane Orientation of A1 Acoustic Delay Lines

The previous analysis assumes A1 ADLs placed along +X (Fig. 1). In this subsection, the effect of in-plane orientation on A1 transduction, propagation, and wideband performance will be discussed.

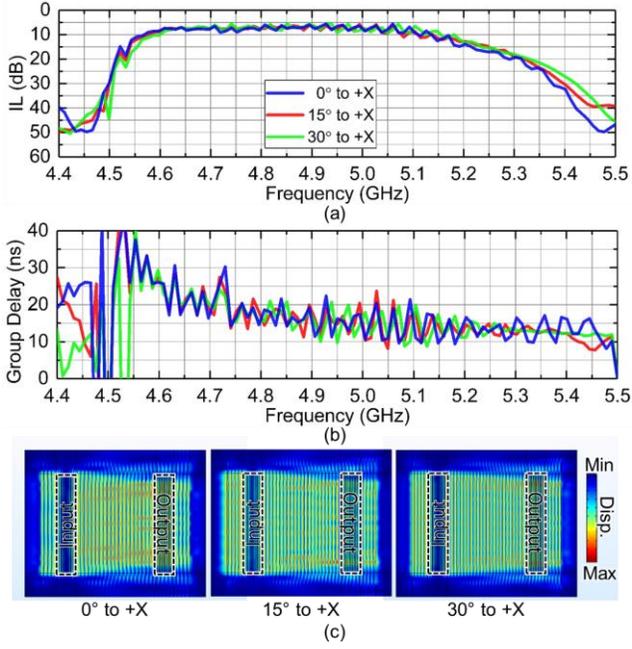

Fig. 15. Simulated effects of the in-plane orientation on ADL performance. (a) IL and (b) group delay of ADLs at different in-plane orientations. (c) Vibration mode shape of A1 in the passband (top view).

A1 characteristics at different in-plane orientations in a Z-cut LiNbO₃ thin film are first investigated. 3D FEA is used to identify the eigenfrequencies of A1 at different orientations, using a 2.4 μm by 50 μm by 0.49 μm Z-cut LiNbO₃ plate. Periodic boundary conditions are applied to the lateral edges. Mechanically free boundary conditions are applied to the top and bottom surfaces. The phase velocities for both electrically open case ($v_f$) and short case ($v_m$) are obtained, respectively. As seen in Fig. 14 (a) and (b), both velocities vary little pertaining to the in-plane orientation. $v_f$ is around 12.52 km/s, and $v_m$ is around 11.70 km/s. A periodicity of 60° is observed in the variation, matching the in-plane angular periodicity of Z-cut LiNbO₃ [74]. $k^2$ is calculated with equation (11) and is plotted in Fig. 14 (c), showing a value (around 14.5%,) in agreement with the calculated in Fig. 3. Clearly from Fig. 14, A1 transduction in Z-cut LiNbO₃ does not vary significantly with the in-plane orientation.

Second, the propagation characteristics of A1 are studied. So far, the analysis assumes that the wavefront propagates in alignment with the energy transportation direction [75]. However, this is only true when the power flow angle (PFA) is zero. The PFA is defined as the in-plane angle between the direction of $v_g$ and $v_p$, pointing from $v_g$ to $v_p$ [75], which is mostly non-zero for waves in anisotropic materials. A large PFA would cause the generated wave propagating off the direction toward the output transducer. Although the free boundaries in the transverse direction would help to confine the energy, IL degradation is still expected as waves scatter into the bus line area where no IDTs are present to collect the acoustic energy [44]. The PFA for A1 waves in Z-cut LiNbO₃ is studied through the slowness curve approach [75] and plotted in Fig. 14 (d) for both the electrically open and short cases. Small PFAs can be seen across the YZ plane. A PFA of 0° is seen along +X. A maximum of +0.6° along 15° to +X and a minimum of −0.6° along 45° to +X are observed. The PFA shows the same periodic dependence on orientation as $v_f$, $v_m$, and $k^2$.

To explore the effects of a small PFA, 3D FEA is set up with a cell length $\Lambda$ of 2.4 μm, a gap length $L_g$ of 40 μm, and a cell number $N$ of 4. The aperture width is 50 μm, and the total device width is 74 μm. PMLs are placed on the longitudinal ends, while the free boundaries are set on the transverse sides. The simulated S parameters are shown in Fig. 15 (a) and (b) with ports conjugately matched to 210 + j140 Ω, showing the minor difference between devices oriented at different angles. The displacement mode shape presented in Fig. 15 (c) shows that most energy propagates along the longitudinal direction. Compared with ADLs using other modes with significant PFAs [42], [44], A1 in Z-cut LiNbO₃ allows more tolerance for angular misalignment due to its small PFAs.

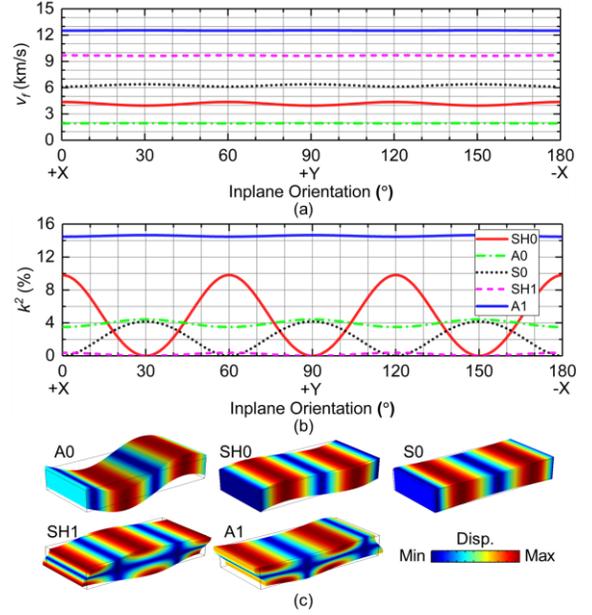

Fig. 16. Simulated characteristics of major modes at different in-plane orientations in a 0.49 μm thick Z-cut LiNbO₃ thin film, including (a) $v_f$, (b) $k^2$, and (c) displacement mode shapes of different modes.

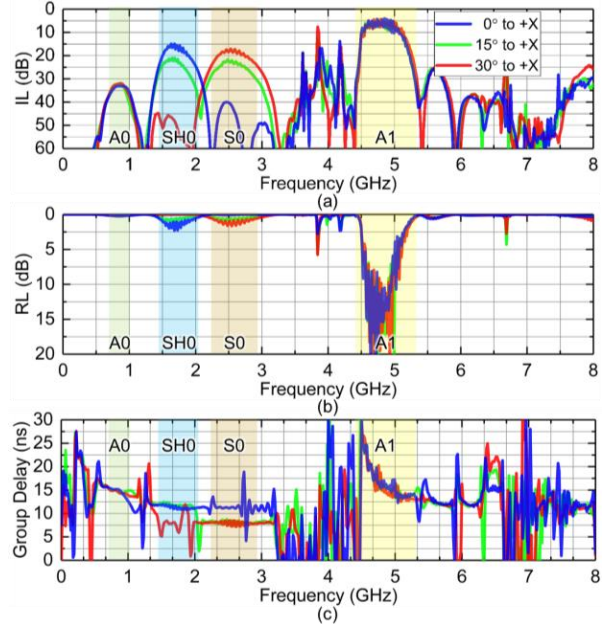

Fig. 17. Simulated wideband (a) IL, (b) RL, and (c) group delay of A1 ADLs at different in-plane orientations.

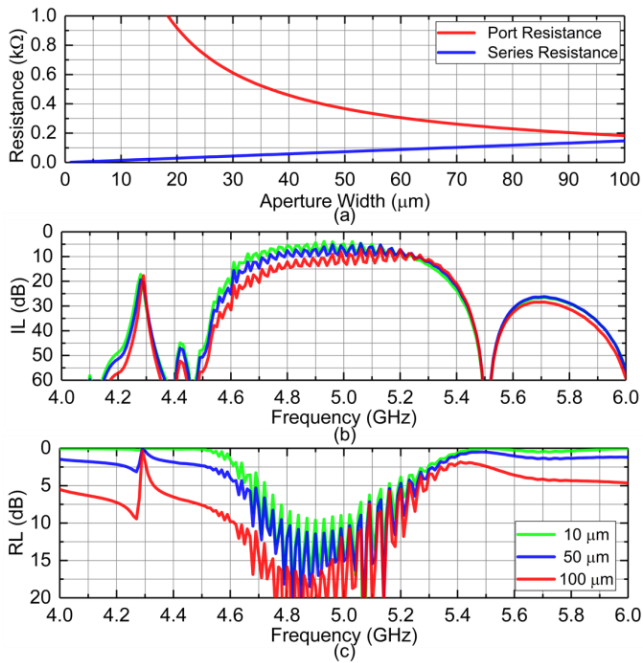

Fig. 18. (a) Simulated effects of the aperture width on port resistance and series resistance. Simulated (b) IL and (c) RL of A1 ADLs with different aperture widths.

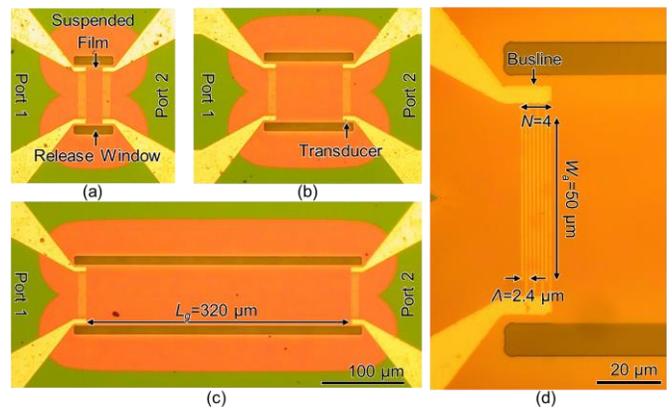

Fig. 19. Optical microscope images of the fabricated ADLs. Zoomed-out images of A1 ADLs with $L_g$ of (a) 20, (b) 80, and (c) 320 μm. (d) Zoomed-in image of a transducer with 4 cells.

TABLE II
KEY PARAMETERS OF THE FABRICATED DEVICES

| Index | Cell Length (μm) | Gap Length (μm) | No. of Cells | Sim. (Fig.) | Meas. (Fig.) | Comments |
|---|---|---|---|---|---|---|
| Group A | 2.4 | 20-320 | 4 | 5, 17 | 20-21 | Gap length & Wideband |
| Group B | 3.2 | 20-320 | 4 | 6 | 22 | Cell length & Gap length |
| Group C | 2.8 | 20-160 | 4 | 6 | 23 | Cell length & Gap length |
| Group D | 2.0 | 20-320 | 4 | 6 | 24 | Cell length & Gap length |
| Group E | 2.4 | 20-320 | 2 | 8 | 25 | Cell number, $v_g$, and PL |

Third, other modes at different angles are studied. $v_f$ and $k^2$ of the major modes in the 2.4 μm by 50 μm by 0.49 μm Z-cut LiNbO₃ plate are simulated using the same method as that for A1. The results are plotted along with the displacement mode shapes in Fig. 16. SH0, S0, and A0 can be effectively excited in Z-cut LN with moderate $k^2$ at certain orientations. The simulated wide-band performance for ADLs placed along 0°, 15°, and 30° to +X is shown in Fig. 17. In agreement with Fig. 16, S0 is not excited at 0° to +X, while SH0 is not excited at 30° to +X. The frequency spacings between passbands mark the difference in $v_p$, while the difference in $\delta$ proves the difference in $v_g$ for different modes. Other than the non-propagating modes below $f_{c\_open}$, a clean spectrum can be observed for A1.

Based on the above analysis of A1 transduction, propagation, and its wideband performance, it can be concluded that the inplane orientation does not affect the performance significantly. Consequently, we will choose the X-axis as the longitudinal direction for device implementation in this work.

### C. Electrical Loading in Interdigitated Electrodes

The series resistance in the IDTs can cause significant performance degradation in a wide device aperture [76]. With a wider aperture (or longer IDTs), the series resistance caused by the electrical loading increases, while the radiation resistance of the ADL decreases. Consequently, the electrical loading effects are more prominent. To study electrical loading quantitatively,

$R_s$ can be calculated as [76]:

$$R_{ele} = (2\rho_s \cdot W_a)/(3t \cdot \Lambda/4) \tag{17}$$

$$R_s = 2R_{ele}/N \tag{18}$$

where $R_{ele}$ is the resistance in a single IDT. $\rho_s$ is the electrical resistivity. $W_a$ is the aperture width, $\Lambda/4$ is the electrode width, and $t$ is the IDT electrode thickness. $R_s$ is the series resistance of a transducer, and $N$ is the cell number. From equations (17)-(18), $R_s$ is proportional to $W_a$. For a device with a $\Lambda$ of 2.4 μm, an $N$ of 4, and 30 nm Al electrodes, $R_s$ can be calculated for different $W_a$ [Fig. 18(a)]. In the calculation, $\rho_s$ is set as 3 times of the bulk value (2.65 × 10⁻⁸ Ω·m [77]), based on that measured from the in-house fabrication tests. The real part of the port impedance (port resistance, $R_{port}$) without considering the electrical loading is inversely proportional to $W_a$, as plotted in Fig. 18 (a). The comparison indicates that the electrical loading is significant for devices wider than 50 μm. To further investigate the impact, the simulated S parameters of ADLs with 10, 50, and 100 μm are shown in Fig. 18 (b) (c), with the port impedance conjugately matched to 1580 + j260 Ω, 420 + j55 Ω, and 295 + j28 Ω respectively. A decrease in IL and an increase in RL are the results of the electrical loading. The impact is more clear on the lower frequency side of the passband because $k^2$ of A1 is slightly larger at lower frequencies. The same $R_s$ is more substantial in comparison to the radiation resistance at those frequencies. Another consequence is that the compound port resistance (when $R_s$ is non-zero) is not inversely proportional to $W_a$ due to the electrical loading. However, it is not beneficial to implement devices with excessively small apertures because of the wave diffraction caused by the fringe effect [7]. Therefore, the aperture width is set as 50 μm as a trade-off. Based on the results, we choose to implement A1 ADLs along the X-axis direction using 30 nm of Al electrode and an aperture width of 50 μm.

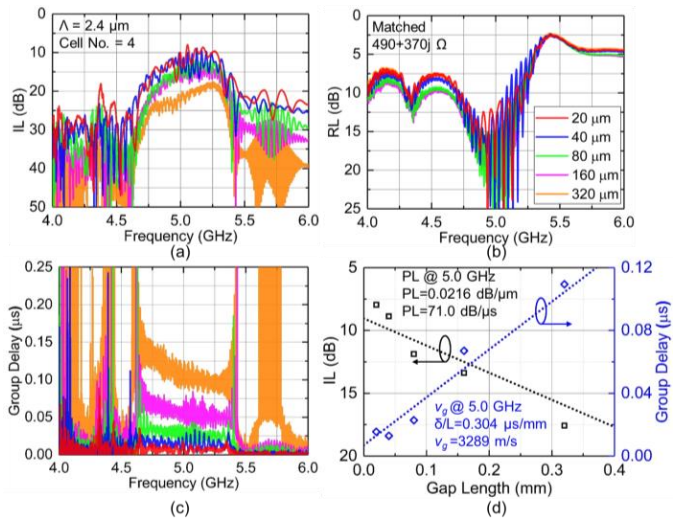

Fig. 20. Measured S-parameters of the A1 ADLs in Group A ($N$=4, $\Lambda$=2.4 μm) with identical transducers but different $L_g$ (20 – 320 μm). (a) IL, (b) RL, and (c) group delay responses. (d) Extracted propagation loss (71 dB/μs), and group velocity (3289 m/s) of A1 at 5.0 GHz.

## IV. A1 ACOUSTIC DELAY LINE IMPLEMENTATION

The devices were fabricated in-house with the process presented in [44]. A 490 nm Z-cut LiNbO₃ thin film on a 4-inch Si wafer is provided by NGK Insulators, Ltd., for the fabrication. The optical images of the fabricated ADLs are shown in Fig. 19. The key design parameters, namely $\Lambda$, $L_g$, and $N$ are labeled, and their typical values are presented in Table I.

Five groups of A1 ADLs are designed for the implementation of 5-GHz broadband delays (Table II). ADLs in group A have the same transducer design ($\Lambda$ and $N$) but different $L_g$, for showcasing the operation principles of A1 ADLs and identifying the key propagation parameters. Their wideband performance will also be presented to validate our design. Groups B, C, and D include ADLs with different cell length $\Lambda$ for showing ADL performance at different frequencies and also present the highly dispersive characteristics of A1. Group E includes ADLs with a different number of cells from Group A to show the dependence of BW on $N$. The broadband performance is also used to extract $v_g$ and PL. Measured results and discussion will be presented in the next section.

## V. MEASUREMENTS AND DISCUSSION

### A. Acoustic Delay Lines with Different Gap Lengths

The fabricated ADLs were first measured with a vector network analyzer (VNA) at the −10 dBm power level in air, and then conjugately matched using ADS. ADLs in Group A ($N = 4$, $\Lambda = 2.4$ μm, $L_g = 20$-320 μm) are designed for showcasing A1 ADL operation and demonstrating long delays.

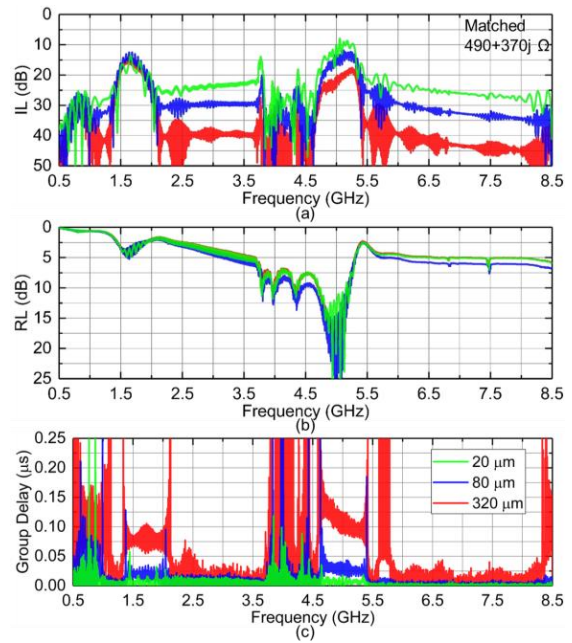

Fig. 21. Measured wideband performance of the devices in Group A: (a) IL, (b) RL, and (c) group delay.

The measured IL and RL are shown in Fig. 20 (a)-(b) with the ports conjugately matched. The ADLs show a passband centered at 5.0 GHz. A minimum IL of 7.9 dB, an average IL of 9.1 dB, and an FBW around 4% have been achieved for the ADL with a 20 μm gap length. The average IL and 3-dB FBW are extracted from the smoothed transmission (1000-point-window from measured data) using the Savitzky-Golay approach [70]. Delays between 15 ns and 109 ns are measured. An increase in IL is observed for longer ADLs, which is caused by the PL of A1 in the LiNbO₃ waveguide. Larger transmission can be observed out of the passband for shorter devices, which is likely caused by the capacitive feedthrough between the buslines and the probing pads. Ripples caused by the multi-reflection between ports and the internal reflections in the transducers are seen in the passband. Larger RL out of the passband is observed, due to the series resistance in the electrodes, as explained in Section III-C. The non-propagating modes can be observed below the cut-off frequency in Fig. 20 (a), but they are significantly damped by PL. Dispersive group delays are observed for different devices, showing longer delays near the cut-off frequency as modeled in Section II-A. A1 propagation characteristics are extracted from the dataset, showing a PL of 71 dB/μs (or 0.0216 dB/μm), and $v_g$ of 3289 m/s 5.0 GHz.

The wideband performance of A1 ADLs is presented in Fig. 21. The cut-off can be clearly identified below the $f_{c\_open}$ around 4.4 GHz where the onset of larger IL occurs. Three out of band resonances are present at 3.7 GHz, 3.9 GHz, and 4.3 GHz, as predicted in Fig. 4. An A0 passband at 0.8 GHz and an SH0 passband at 1.6 GHz are also measured, consistent with simulations in Fig. 17. Different group delays are observed in the A1 and SH0 passbands as A1 is slower than SH0 in this frequency range. This validates that A1 features low $v_g$ and high $v_p$ simultaneously, promising compact device sizes while maintaining large feature sizes at 5 GHz.

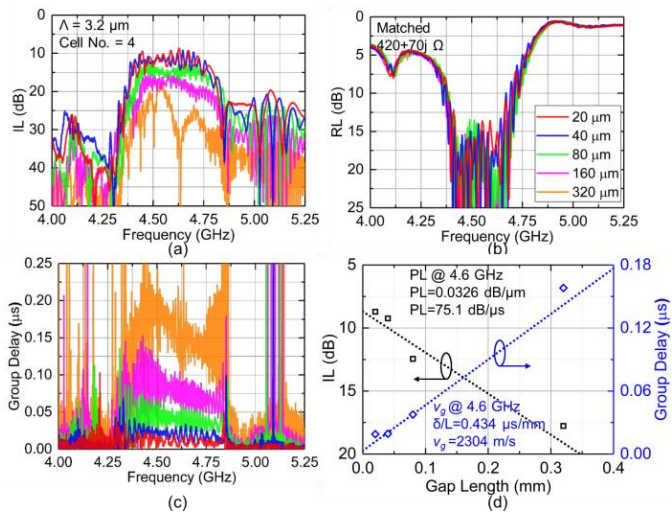

Fig. 22. Measured S-parameters of the A1 ADLs in Group B ($N$=4, $\Lambda$=3.2 μm) with identical transducers but different $L_g$ (20 – 320 μm). (a) IL, (b) RL, and (c) group delay responses. (d) Extracted propagation loss (75.1 dB/μs), and group velocity (2304 m/s) of A1 at 4.6 GHz.

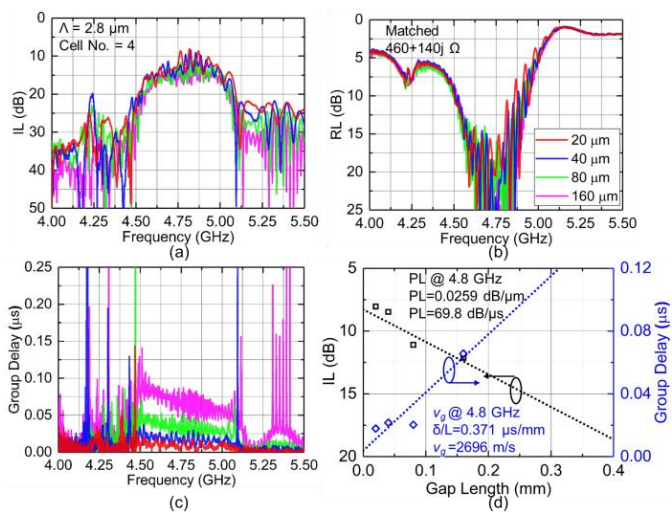

Fig. 23. Measured S-parameters of the A1 ADLs in Group C ($N$=4, $\Lambda$=2.8 μm) with identical transducers but different $L_g$ (20 – 160 μm). (a) IL, (b) RL, and (c) group delay responses. (d) Extracted propagation loss (69.8 dB/μs), and group velocity (2696 m/s) of A1 at 4.8 GHz.

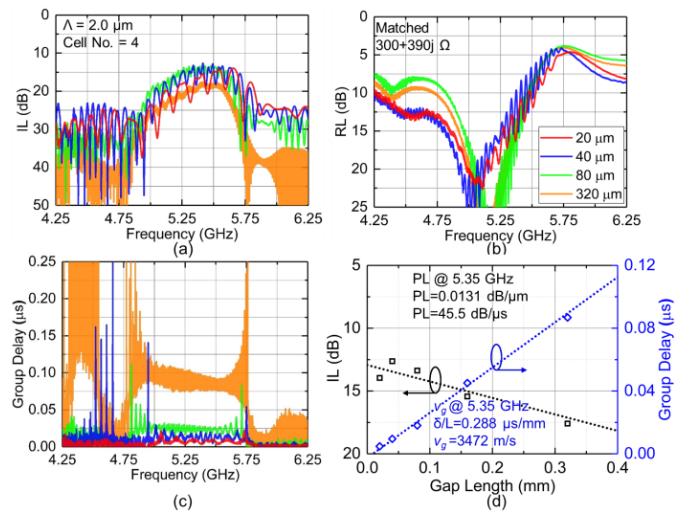

Fig. 24. Measured S-parameters of the A1 ADLs in Group D ($N$=4, $\Lambda$=2.0 μm) with identical transducers but different $L_g$ (20 – 320 μm). (a) IL, (b) RL, and (c) group delay responses. (d) Extracted propagation loss (45.5 dB/μs), and group velocity (3472 m/s) of A1 at 5.35 GHz.

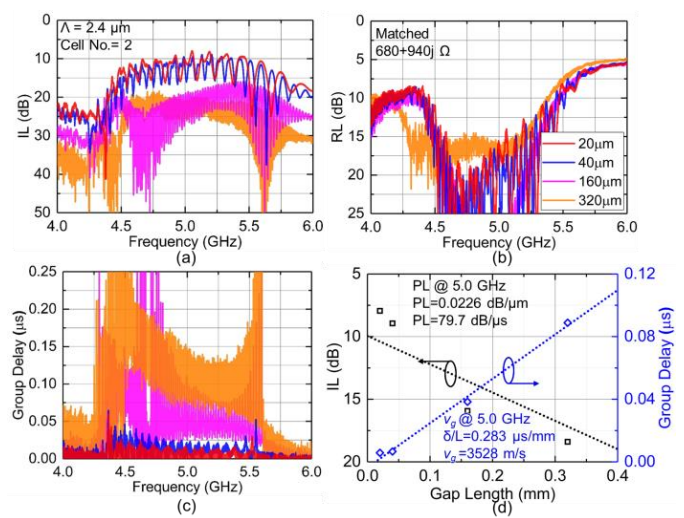

Fig. 25. Measured S-parameters of the A1 ADLs in Group E ($N$=2, $\Lambda$=2.4 μm) with identical transducers but different $L_g$ (20 – 320 μm). (a) IL, (b) RL, and (c) group delay responses. (d) Extracted propagation loss (79.7 dB/μs), and group velocity (3528 m/s) of A1 at 5 GHz.

### B. Acoustic Delay Lines with Different Center Frequencies

ADLs in Group B ($\Lambda$ = 3.2 μm), Group C ($\Lambda$ = 2.8 μm), and Group D ($\Lambda$ = 2.0 μm) are designed for investigating the impact of the cell length on the center frequency. In each group, devices with gap length between 20 and 320 μm are implemented.

Devices are measured at -10 dBm in air and conjugately matched. For devices in Group B, a minimum IL of 8.71 dB, an average IL of 10.4 dB, and an FBW around 5.7%, and a center frequency of 4.6 GHz are obtained (Fig. 22). The extracted PL is 75.1 dB/μs (or 0.0326 dB/μm), and $v_g$ of 2304 m/s for A1 at 4.6 GHz. For devices in Group C, a minimum IL of 8.04 dB, an average IL of 11.2 dB, and an FBW around 7.3%, and a center frequency of 4.8 GHz are obtained (Fig. 23). The extracted PL is 69.8 dB/μs (or 0.0259 dB/μm), and $v_g$ of 2696 m/s for A1 at 4.8 GHz. For devices in Group D, a minimum IL of 12.6 dB, an average IL of 14.5 dB, and an FBW around 8.8%, and a center frequency of 5.35 GHz are measured (Fig. 24). The extracted PL is 45.5 dB/μs (or 0.0131 dB/μm), and $v_g$ of 3472 m/s for A1 at 5.35 GHz. Comparing the performance between ADLs from different groups, devices with larger cell lengths have lower center frequencies. However, unlike S0 and SH0 [42], [44], the A1 center frequency does not scale inversely to the cell length due to the dispersive nature of A1. Moreover, higher frequency devices tend to have flatter group delays in the passband, which are consistent with Fig. 3.

### C. Acoustic Delay Lines with Different Cell Numbers

ADLs in Group E ($N$ = 2, $\Lambda$ = 2.4 μm, $L_g$ = 20-320 μm) are designed for investigating the impact of cell numbers on the bandwidth via comparison with Group A. The passband is not symmetrical with larger IL shown below $f_{c\_open}$ due to the cut-off. For devices in Group E, a minimum IL of 7.9 dB, an

### TABLE III
### EXTRACTED A1 MODE PROPAGATION CHARACTERISTICS

| Index | $f_{center}$ (GHz) | Group Velocity | | Delay/length (µs/mm) | PL | |
|---|---|---|---|---|---|---|
| | | $v_g$ (m/s) | | | PL/length (dB/µm) | PL/delay (dB/µs) |
| Group A | 5.0 | 3289 | | 0.304 | 0.0216 | 71.0 |
| Group B | 4.6 | 2304 | | 0.434 | 0.0326 | 75.1 |
| Group C | 4.8 | 2696 | | 0.371 | 0.0259 | 69.8 |
| Group D | 5.35 | 3472 | | 0.288 | 0.0131 | 45.5 |
| Group E | 5.0 | 3528 | | 0.283 | 0.0226 | 79.7 |

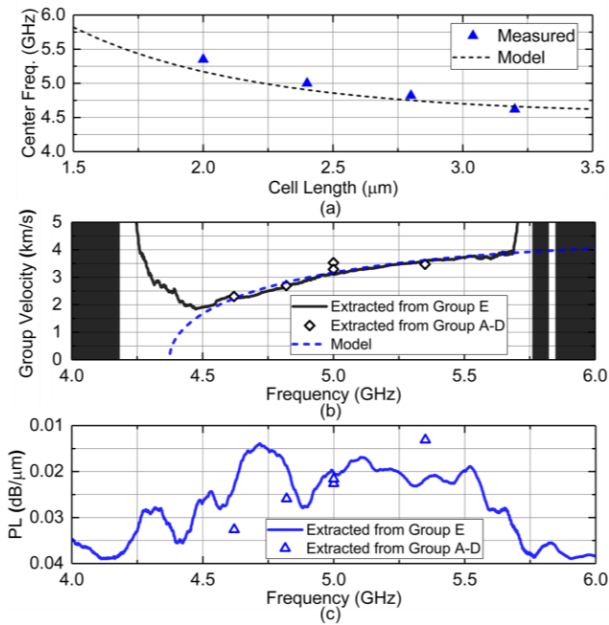

Fig. 26. Extracted parameters of A1 ADLs. (a) Center frequencies of different devices in comparison to the calculated values. (b) Wideband group delay extracted from Groups A-E in comparison with FEA (Fig. 3). (c) Extracted PL from Groups A-E.

average IL of 10.7 dB, and an FBW around 19.7%, and a center frequency of 5.0 GHz are obtained (Fig. 23). The extracted PL is 79.73 dB/µs (or 0.0226 dB/µm), and $v_g$ is 3528 m/s at 4.8 GHz. The data in Group E will be used to extract the wideband $PL$ and $v_g$ for A1 ADLs.

### D. Performance Summary and Discussion

The extracted propagation parameters of different ADLs are presented in Table III, and plotted in Fig. 26. First, the center frequencies $f_{center}$ of different groups are plotted in Fig. 26 (a), and compared to that calculated using the approach in Section III-A (Fig. 12). Good agreement is obtained between the measurement and the model, with the slight differences likely caused by the approximations in the model. Second, the extracted group velocity is presented in Fig. 26 (b). The values obtained from the center frequencies of different groups are plotted using the scattered points. The wideband performance obtained from Group E is also extracted using least square fitting [71] at each frequency point in Fig. 25 (c). The FEA results (Fig. 3) are also plotted on the same figure, showing great agreement with measured data. The extracted group velocity validates the cut-off. Finally, PL at different frequencies is plotted in Fig. 26 (c). Similarly, PL from different groups and the wideband PL from Group E are plotted. The extracted PL of A1 in thin-film LiNbO₃ around 5 GHz is reported for the first time. Interestingly, smaller PL per distance is observed at higher frequencies. The reason has not been identified and will be investigated in future studies where the passband ripples are suppressed through unidirectional transducers [34].

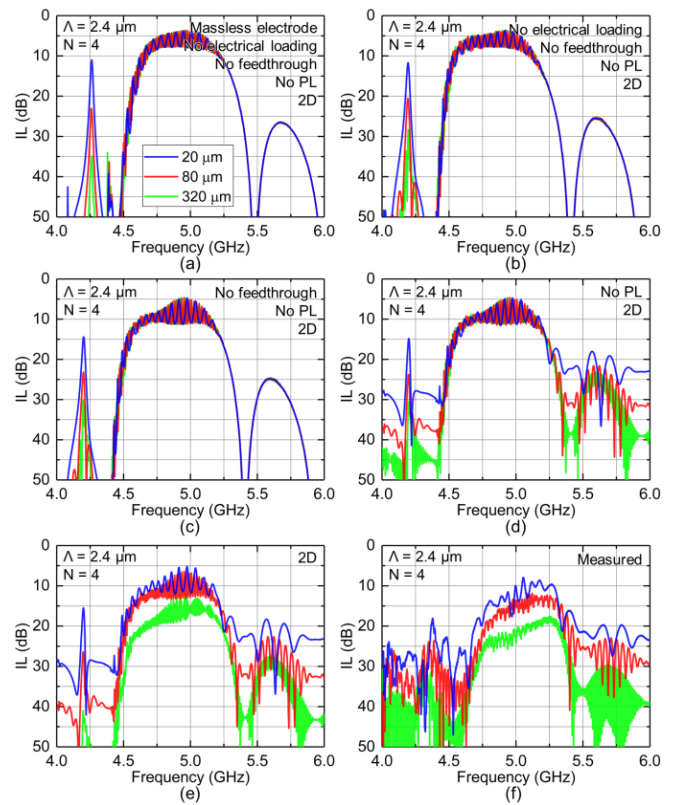

Fig. 27. Comparison between the simulated and measured IL of the devices in Group A. Different practical considerations are gradually included, based on (a) the ideal A1 ADL, including (b) electrode mass loading, (c) electrical loading, (d) feed through capacitance, and (e) propagation loss. Simulations are compared with the measured results in (f).

Finally, to identify the major loss contributors, the measured performance is compared to the simulated values discussed in Sections II and III. Without loss of generality, the IL of ADLs in Group A with $L_g$ of 20, 80, and 320 µm (under conjugate matching conditions) are analyzed in Fig. 27. The 2D FEA of an ideal A1 ADL with massless electrodes, no electrical or mechanical loss, without considering direct capacitance feedthrough between probing pads is presented in Fig. 27 (a). The mass loading is first introduced [Fig. 27 (b)]. Afterward, the electrical loading is included [Fig. 27 (c)], introducing additional IL in the passband (Section III-C). The other element influencing the performance is the capacitive feedthrough between the pads and buslines (Section V-A). Fig. 27 (d) is plotted with 3.2 fF for the 20-µm $L_g$ device, 1.2 fF for the 80-µm $L_g$ device, and 0.5 fF for the 320-µm $L_g$ device, where the capacitance is fitted from the measured out-of-band-rejection. The final consideration is the experimentally-measured PL of A1 (Section V-D), which causes additional in-band IL [Fig. 27 (e)].

TABLE IV REPORTED ADL PERFORMANCE WITH SUB-20 DB IL

| Reference | IL (dB)* | FBW (%)** | $f_{center}$ (GHz)** |
|---|---|---|---|
| [34] | 7.2 | 1.0% | 0.19 |
| [18] | 10 | 1.0% | 0.28 |
| [80] | 4.0 | 0.5% | 0.25 |
| [81] | 9.7 | 2.0% | 0.4 |
| [82] | 18 | 8.5% | 2.1 |
| [31] | 5.5 | 3.8% | 2.4 |
| [46] | 1.0 | 4.0% | 0.30 |
| [42] | 3.7 | 4.0% | 0.97 |
| [44] | 3.2 | 3.9% | 0.96 |
| This work | 7.9 | 4.0% | 5.0 |

*Defined by minimum in-band IL. **Values extracted from the figures in the reports.

The average IL is 8.7 dB, and the FBW is 10%. The simulated value achieves reasonable agreement with the measurement (an average IL of 9.1 dB and an FBW of 4%) in Fig. 27 (f). The difference in FBW is caused by the higher PL at lower frequencies. The additional IL in the measurement likely originates from the wave diffraction and the additional electrical loss in the IDTs at higher frequencies.

Another issue of the A1 ADL prototypes is the relatively high port impedance. To improve in the future, several approaches can be taken. First, devices with larger aperture widths can lower the port resistance. However, it is only feasible after reducing the electrical resistivity in the IDTs (Fig. 18) with better controlled sputtering conditions and annealing processes [78], [79]. Second, more transducer pairs can be adopted to lower the port impedance at the cost of a smaller FBW (Section II-D), provided device performance can still meet the application specifications.

To sum up, we have demonstrated A1 ADLs at 5 GHz in LiNbO3 thin films for the first time, significantly surpassing the operation frequencies of the previous reports (Table IV). Note that the demonstrated device performance is still far from the full potential of A1 ADLs. Upon further optimizations, lower IL devices with less pronounced passband ripples and matched to a lower port impedance can be expected.

## VI. CONCLUSION

In this work, we have demonstrated A1 ADLs at 5 GHz in LiNbO$_3$ thin films for the first time. Thanks to the fast phase velocity, significant coupling coefficient, and low-loss of A1, the demonstrated ADLs significantly surpass the state of the art with similar feature sizes in center frequency. The propagation characteristics of A1 in LiNbO$_3$ are analyzed and modeled with FEA before the designs of A1 ADLs are studied and composed. The implemented ADLs at 5 GHz show a minimum IL of 7.9 dB, an average IL of 9.1 dB, and a 3 dB FBW around 4%. The design variations show delays ranging between 15 ns and 109 ns and the center frequencies between 4.5 GHz and 5.25 GHz. From these measured devices, the propagation characteristics of A1 are extracted for the first time and shown to match our analysis. Upon further optimization, the A1 ADLs can lead to wide-band and high-frequency signal processing functions for 5G applications.

## ACKNOWLEDGMENTS

The authors would like to thank Dr. Troy Olsson for helpful discussions.